\documentclass[aps,prl,superscriptaddress,showpacs,showkeys,a4paper,10pt,notitlepage,twocolumn]{revtex4-1}
\usepackage[utf8]{inputenc}
\usepackage[english]{babel}
\usepackage{amsmath}
\usepackage{amssymb}
\usepackage{amsfonts}
\usepackage{graphicx}
\usepackage{cancel}
\usepackage{upgreek}
\usepackage{mathtools}
\usepackage[T1]{fontenc}

\usepackage{xcolor}
\definecolor{link_blue}{RGB}{52,46,157}

\usepackage{longtable}
\usepackage[colorlinks,citecolor=link_blue,linkcolor=link_blue,urlcolor=link_blue]{hyperref}
\usepackage[tiny,center,uppercase]{titlesec}

\renewcommand{\vec}{\boldsymbol}

\newcommand{\opA}[1]{\boldsymbol{\mathsf{#1}}}
\newcommand{\opa}[1]{\mathsf{#1}}
\newcommand\ri{\mathrm{i}}

\begin{document}

\title{Exact solution for the quantum Rabi model with the
  $\opA{A}^{2}$ term}

\author{I.\ D.\ Feranchuk}
% \email[Corresponding author: ]{ilya.feranchuk@tdt.edu.vn}
\affiliation{Atomic Molecular and Optical Physics Research Group,
  Advanced Institute of Materials Science, Ton Duc Thang University,
  Ho Chi Minh City, Vietnam}
\affiliation{Faculty of Applied Sciences, Ton Duc Thang University, Ho
  Chi Minh City, Vietnam}

\author{N.\ Q.\ San}
\affiliation{Belarusian State University, 4 Nezavisimosty Ave.,
  220030, Minsk, Belarus}

\author{A.\ U.\ Leonau}
\affiliation{Belarusian State University, 4 Nezavisimosty Ave.,
  220030, Minsk, Belarus}

\author{O.\ D.\ Skoromnik}
\email[Corresponding author: ]{olegskor@gmail.com}
\affiliation{Ho Chi Minh City University of Education, 280 An Duong
  Vuong, District 5, Ho Chi Minh City, Vietnam}

\begin{abstract}
  Quantum Rabi model (QRM) is widely used for the analysis of the
  radiation-matter interaction at the fundamental level in cavity
  quantum electrodynamics. Typically the QRM Hamiltonian includes only
  $\opA{p} \cdot \opA{A}$ term, however, the complete nonrelativistic
  Hamiltonian of quantum electrodynamics includes $\opA{A}^{2}$ term
  as well. Here we find an exact solution and demonstrate with the
  help of the exact canonical transformations that the QRM Hamiltonian
  with the $\opA{A}^{2}$ term (QRMA) is reduced to the standard QRM
  model Hamiltonian with the renormalized frequency and the coupling
  constant and the eigenstates are expressed through the squeezed
  states of the field. As a result, the $\opA{A}^{2}$ term
  qualitatively changes the behavior of the QRM with purely
  electromagnetic interaction in the strong coupling regime: the value
  of the ground state energy of an atom inside the cavity is higher
  than in vacuum and the number of crossing of energy levels with
  different quantum numbers decreases.
\end{abstract}

\keywords{}
\maketitle

\textbf{After preparing the work for publication the authors became
  aware of the recent work \cite{di_stefano_resolution_2019} where the
  derivation of the Hamiltonian of a two level system is provided and
  was shown that the truncation of the atomic Hilbert space to the two
  levels leads to an incorrect Hamiltonian with the $\opA{A}^{2}$ term
  (\ref{eq:2}).}

\textbf{As a result all conclusions of our manuscript are based on the
  form of the Hamiltonian (\ref{eq:2}), which does not describe a two
  level system. If by some other means the Hamiltonian (\ref{eq:2})
  can be engineered, e.g. in a cold atoms system or by any other mean,
  the conclusions will be valid for that system.}

\textbf{The easiest way to demonstrate the equivalence of the
  Hamiltonian of a two level system with the $\opA{A}^{2}$ term and
  the Hamiltonian with the $\opA{E}\cdot \vec r$ is provided in the
  Ref.~\cite{scully_quantum_1997} Eqs.~(5.1.12-5.1.17).}

\textbf{Concluding, the Hamiltonian of the two level system with the
  $\opA{A}^{2}$ term in the dipole approximation is equivalent to the
  following Hamiltonian}
\begin{align}
  \opa{H} = \frac{\opA{p}^{2}}{2m_{0}} + \opa{V}(r) - e_{0}
  \opA{E}\cdot\vec{r}. \label{eq:1} 
\end{align}

Quantum Rabi model describes the interaction of a two level atom with
a single-mode quantum field in a cavity
\cite{PhysRev.49.324, PhysRev.51.652}. This model plays fundamental
role in the radiation-matter interaction in cavity quantum
electrodynamics \cite{PhysRevA.99.033823, PhysRevA.99.033834,
  PhysRevLett.122.123604}, quantum optics \cite{Walther_2006}, quantum
information \cite{RevModPhys.73.565} and physics of condensed matter
\cite{holstein_studies_1959}. In addition, it was recently demonstrated
\cite{PhysRevLett.107.100401} that the QRM is an exactly integrable
system and the problem of determining its spectrum of stationary
states is reduced to the numerical solution of many-term recurrent
relations. Furthermore, many approximate methods were developed for
the description of QRM, among which, the most widely used is the
rotating wave approximation (RWA) applicable for small values of the
detuning of the field frequency from the resonant atomic transition
and small values of the coupling constant $f$ of the atom-field
interaction. The RWA approximation is based on the exact solution of
the Schr\"odinger equation with the Jaymes-Cummings Hamiltonian
\cite{jaynes_comparison_1963}.

Presently there exist systems (ion bound to the cavity
\cite{PhysRevLett.124.013602, forn-diaz_ultrastrong_2017},
super-conducting qubit \cite{forn-diaz_broken_2016,
  PhysRevLett.115.133601, yoshihara_superconducting_2017}, polaritons
\cite{PhysRevLett.119.127401}) that provide the strong interaction
\cite{murch_beyond_2017, PhysRevA.80.032109, niemczyk_circuit_2010,
  PhysRevLett.105.023601, PhysRevLett.109.193602} between an atom and
the field, which corresponds to the situation when the dimensionless
coupling constant $f$ of the atom-field interaction in QRM reaches
values of the order of unity \cite{RevModPhys.91.025005,
  PhysRevB.72.115303, xie_quantum_2017}. This motivates both
experimental and theoretical investigations of physical effects
appearing in this fully quantum regime \cite{RevModPhys.91.025005,
  PhysRevA.99.033834, feranchuk_physical_2016, saiko_relaxation_2014,
  PhysRevA.78.053805, feranchuk_spontaneous_2017,
  feranchuk_resonant_2009}.

In its conventional form the QRM includes only $\opA{p} \cdot \opA{A}$
term in the Hamiltonian. However, the exact Hamiltonian of
nonrelativistic quantum electrodynamics includes the $\opA{A}^{2}$
term as well. Consequently, the question arises how the system
behaviour in the strong-coupling regime is modified by the inclusion
of $\opA{A}^{2}$ term in the Hamiltonian of the QRM, i.e QRM with the
$\opA{A}^{2}$ (QRMA). In this letter we employ the exact canonical
transformations of the field variables and demonstrate that the
Hamiltonian of QRMA is reduced to the standard QRM Hamiltonian with
the renormalized frequency $\Omega = \sqrt{1 + af^{2}}$, the coupling
constant $\tilde{f} = f/(1+a f^{2})^{1/4}$ with $a>0$ and a constant
energy shift. The appearance of the constant energy shift together
with the form of the renormalized coupling constant $\tilde{f}$
qualitatively modify the system's behavior and allow effectively the
description of QRMA within the RWA with rather high accuracy (see
below). We point out here that a similar situation arises for the
Dicke model \cite{PhysRevLett.35.432, PhysRevA.97.043820} and a
harmonic oscillator interacting with a quantum field
\cite{PhysRevA.44.563} where the inclusion of the $\opA{A}^{2}$
changes the behavior of the system.

In order to perform the canonical transformations of the field
variables and demonstrate the reduction of the QRMA Hamiltonian to the
QRM one with the renormalized frequency and the coupling constant we
start with the Hamiltonian of a nonrelativistic atom, which interacts
with a single-mode quantum electromagnetic field
\cite{scully_quantum_1997}
$\opa{H}_{\mathrm{a}} = (\opA{p} - e_{0}\opA{A}(\vec r))^{2}/2m_{0} +
\opa{V}(\vec r) + \opA{E}^{2}(\vec r) - \opA{B}^{2}(\vec r)$ in the
dipole approximation. Here $e_{0}$ and $m_{0}$ are the electron charge
and mass respectively, $\opa{V}(\vec r)$ is the binding potential of
an interaction between an electron and a nucleus. If an atom is
approximated only with two levels $\chi_{\uparrow}$,
$\chi_{\downarrow}$ and the transition atomic frequency
$\omega_{\uparrow\downarrow}$ is almost coincident with the field
frequency $\omega$, $\omega_{\uparrow\downarrow} \approx \omega$ then
the Hamiltonian $\opa{H}_{\mathrm{a}}$ of an atom is reduced to the
Hamiltonian of the QRMA
($\opa{H}_{\mathrm{a}} \equiv \opa{H}_{\mathrm{QRMA}}$), which in the
dimensionless variables $\hbar = c = 1$ reads
\cite{scully_quantum_1997}
\begin{align}
  \opa{H}_{\mathrm{QRMA}} = \frac{\Delta}{2} \sigma_{3} + f(\opa{a} +
  \opa{a}^{\dag})\sigma_{1} + \opa{a}^{\dag}\opa{a} + k (\opa{a} +
  \opa{a}^{\dag})^{2}, \label{eq:2}
\end{align}
where $f = e_{0} \omega \Delta d \sqrt{2\pi / (\omega^{3} V)}$,
$k = 2\pi e_{0}^{2} / (2m_{0} \omega^{2} V)$, $V$ is the volume of a
cavity, $\Delta = \omega_{\uparrow\downarrow} / \omega$ is the
resonant atomic frequency measured in the units of the electromagnetic
field frequency $\omega$, $\opa{a}$ and $\opa{a}^{\dag}$ are the
annihilation and creation operators of the quantum field,
$[\opa{a}, \opa{a}^{\dag}] = 1$ and $d$ is the dipole matrix element of the
transition between atomic states $\chi_{\uparrow}$ and
$\chi_{\downarrow}$.

Let us now introduce a unitary operator
\begin{align}
  \opa{S} = \exp\left(\frac{1}{4} (\opa{a}^{2} - \opa{a}^{\dag2})
  \ln\Omega\right), \quad \opa{S}^{\dag} = \opa{S}^{-1}\label{eq:3}
\end{align}
with a free parameter $\Omega$, which will be determined later. The
operator $\opa{S}$ transforms the field operators $\opa{a}$ and
$\opa{a}^{\dag}$ as \cite{feranchuk_non-perturbative_2015}
\begin{equation}
  \begin{aligned}
    \opa{S}^{\dag}\opa{a}\opa{S}
    &= \frac{1}{2} \left[
      \left(\frac{1}{\sqrt{\Omega}} + \sqrt{\Omega}\right) \opa{a} +
      \left(\frac{1}{\sqrt{\Omega}} - \sqrt{\Omega}\right) \opa{a}^{\dag}
    \right],
    \\
    \opa{S}^{\dag}\opa{a}^{\dag}\opa{S}
    &= \frac{1}{2} \left[ \left(\frac{1}{\sqrt{\Omega}} +
        \sqrt{\Omega}\right) \opa{a}^{\dag} +
      \left(\frac{1}{\sqrt{\Omega}} - \sqrt{\Omega} \right) \opa{a}
    \right],
  \end{aligned}\label{eq:4}
\end{equation}
which corresponds to the introduction of a new vacuum state
$|\Omega\rangle = \opa{S}|0\rangle$ of an electromagnetic field
in a form of a squeezed state \cite{scully_quantum_1997}.
\begin{figure}[t]
  \centering
  \includegraphics[width=\columnwidth]{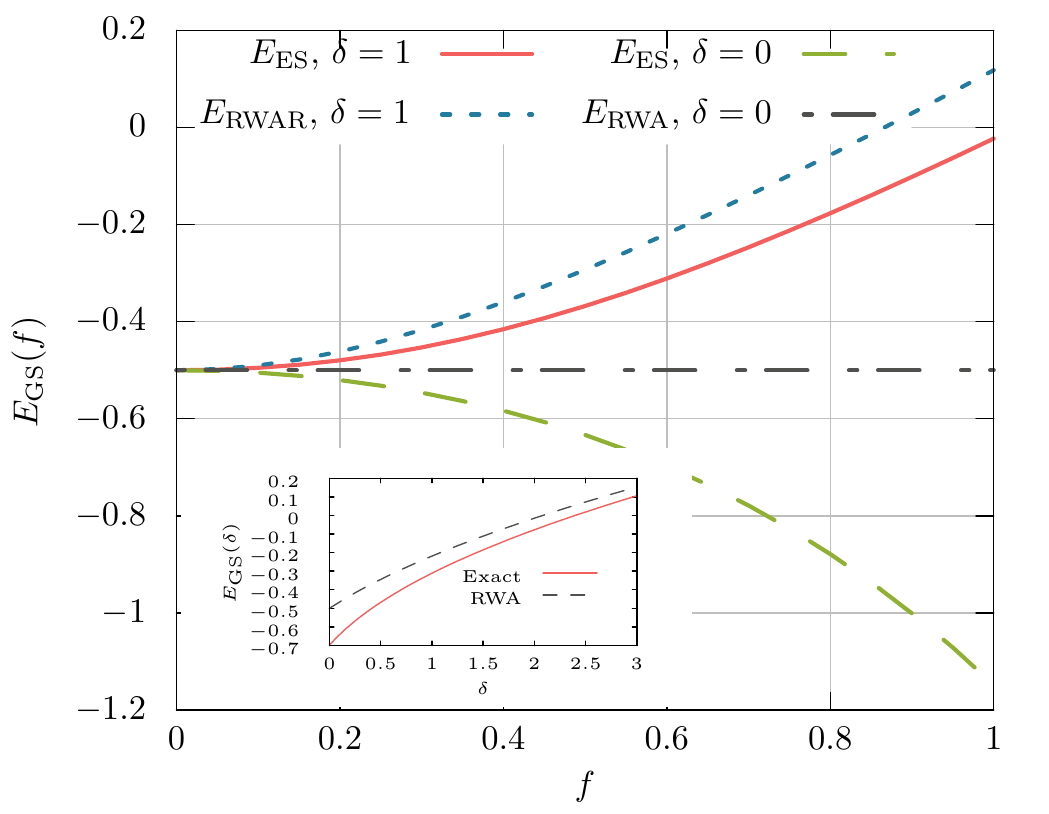}
  \caption{(Color online) The energy of the ground state
    $E_{\mathrm{GS}}$ as a function of the dimensionless coupling
    $f$. We compare the exact numerical solution $E_{\mathrm{ES}}$
    with the rotating wave approximation $E_{\mathrm{RWA}}$ for two
    cases. The first case corresponds to the conventional Rabi model
    with only $\opA{p}\cdot\opA{A}$ term, i.e., $\delta = 0$
    ($E_{\mathrm{RWA}}$). The second case corresponds to the quantum
    Rabi model with the $\opA{A}^{2}$ term included, i.e.,
    $\delta = 1$ ($E_{\mathrm{RWAR}}$). The parameter $\Delta =
    1$. The inset describes the dependence of the ground state energy
    on the parameter $\delta$ and compares the exact solution and RWAR
    for the coupling constant $f = 0.6$.}\label{fig:1}
\end{figure}

As a result of the transformation with the squeezed state operator
$\opa{S}$  the Hamiltonian of the system changes to
\begin{align}
  \opa{H}' = \opa S^{\dag} \opa H_{\mathrm{QRMA}} \opa S
  &= \frac{\Delta}{2}
  \sigma_{3} + \frac{f}{\sqrt{\Omega}} (\opa{a} + \opa{a}^{\dag})
    \sigma_{1} \label{eq:5}
  \\
  &+\frac{k}{\Omega} (\opa{a}^{2} + \opa{a}^{\dag2} + 2 \opa{a}^{\dag}
    \opa{a} + 1) \nonumber
  \\
  &+\frac{1}{4 \Omega} \Bigg[(1 - \Omega^2) (\opa{a}^{2} +
    \opa{a}^{\dag2}) \nonumber
  \\
  &\mspace{40mu}+ 2 (1+\Omega^2)\opa{a}^{\dag}\opa{a} +
    (1-\Omega)^2\Bigg]. \nonumber
\end{align}
\begin{figure}[t]
  \centering
  \includegraphics[width=\columnwidth]{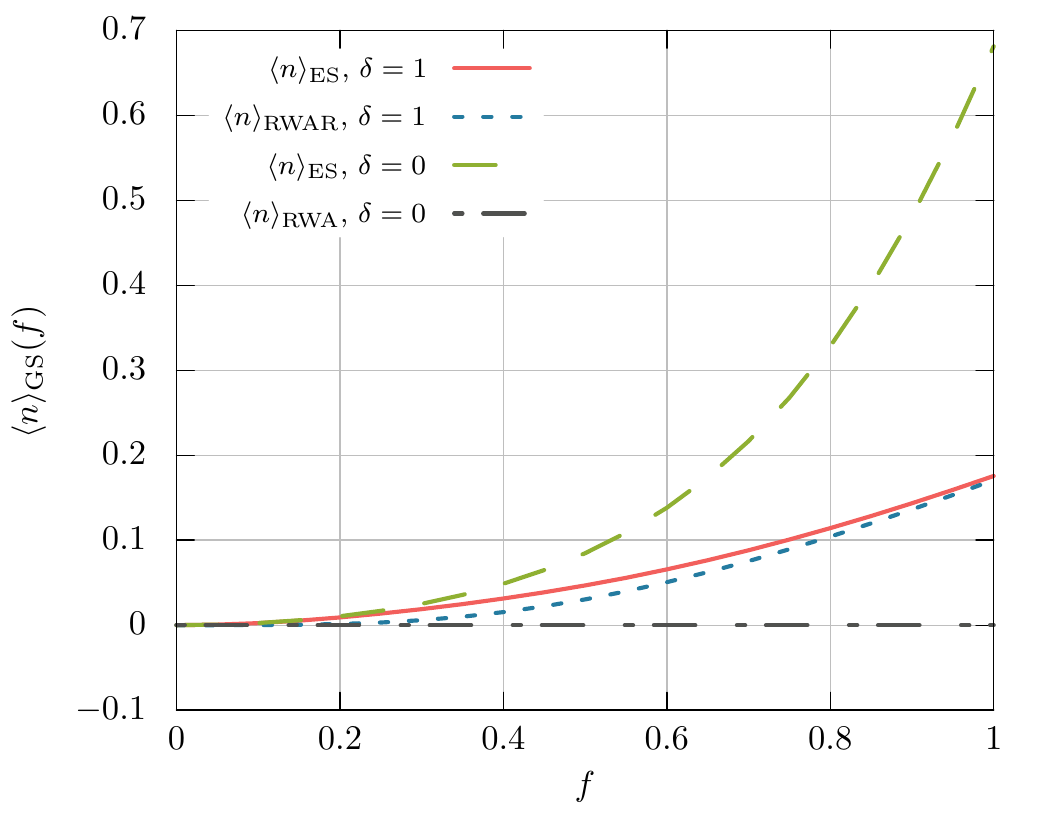}
  \caption{(Color online) The expectation value of the photon number
    operator
    $\langle n\rangle = \langle
    \psi_{\mathrm{GS}}|\opa{a}^{\dag}\opa{a}|\psi_{\mathrm{GS}}\rangle$
    as a function of the dimensionless coupling $f$. We compare the
    exact numerical solution $\langle n\rangle_{\mathrm{ES}}$ with the
    rotating wave approximation $\langle n\rangle_{\mathrm{RWA}}$ for
    the two cases. The first case corresponds to the conventional Rabi
    model with only $\opA{p}\cdot\opA{A}$ term, i.e., $\delta = 0$
    ($\langle n\rangle_{\mathrm{RWA}}$). The second case corresponds
    to the quantum Rabi model with the $\opA{A}^{2}$ term included,
    i.e., $\delta = 1$ ($\langle n\rangle_{\mathrm{RWAR}}$). The
    parameter $\Delta = 1$.}\label{fig:2}
\end{figure}

Now we choose the parameter $\Omega$ from the condition that the
quadratic terms of creation and annihilation operators vanish
\begin{align}
  \Omega = \sqrt{1 + 4k} \label{eq:6}
\end{align}
that leads to the transformed Hamiltonian of the QRMA
\begin{align}
  \opa{H}' = \frac{\Delta}{2} \sigma_{3} + \frac{f}{\sqrt{\Omega}}
  (\opa{a} + \opa{a}^{\dag}) \sigma_1 + \Omega \opa{a}^{\dag}\opa{a} +
  \frac{\Omega - 1}{2}, \label{eq:7}
\end{align}
which is indeed the Hamiltonian of the QRM with the renormalized
frequency and the coupling constant. In addition, there also exists a
constant energy shift. This transformation demonstrates that the QRMA
in a full analogy with the QRM is an exactly integrable system
\cite{PhysRevLett.107.100401}.
\begin{figure*}[t]
  \includegraphics[width=\columnwidth]{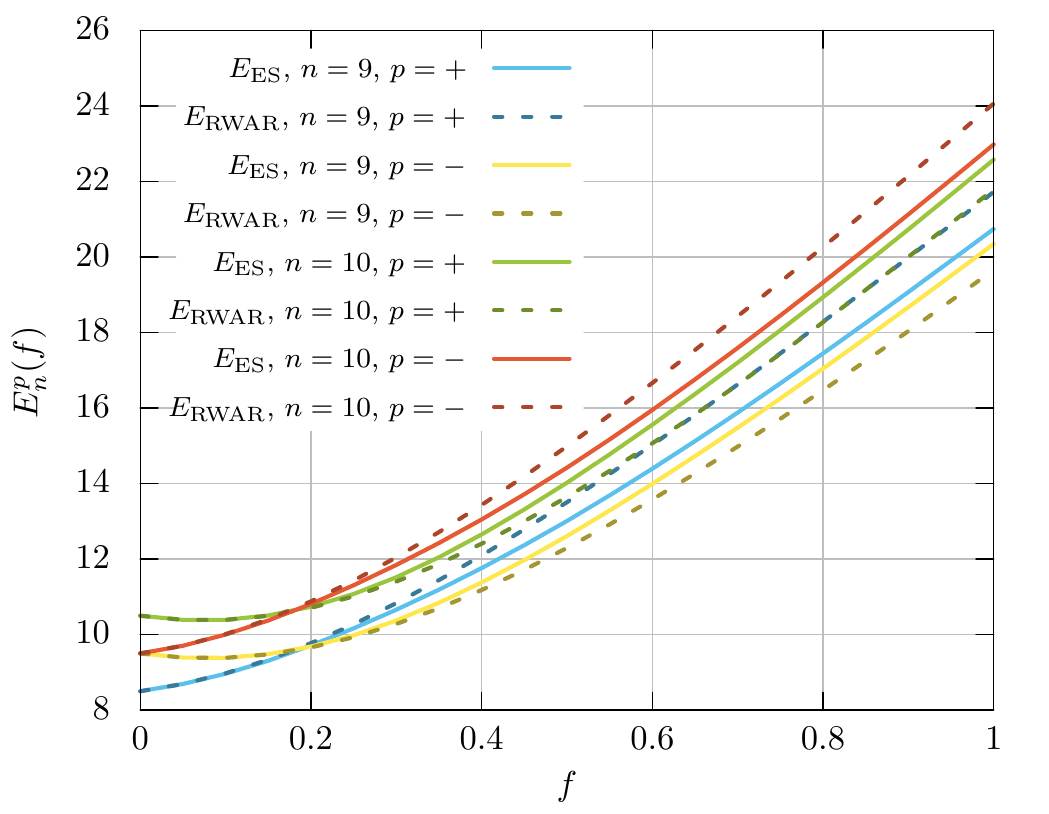}
  \includegraphics[width=\columnwidth]{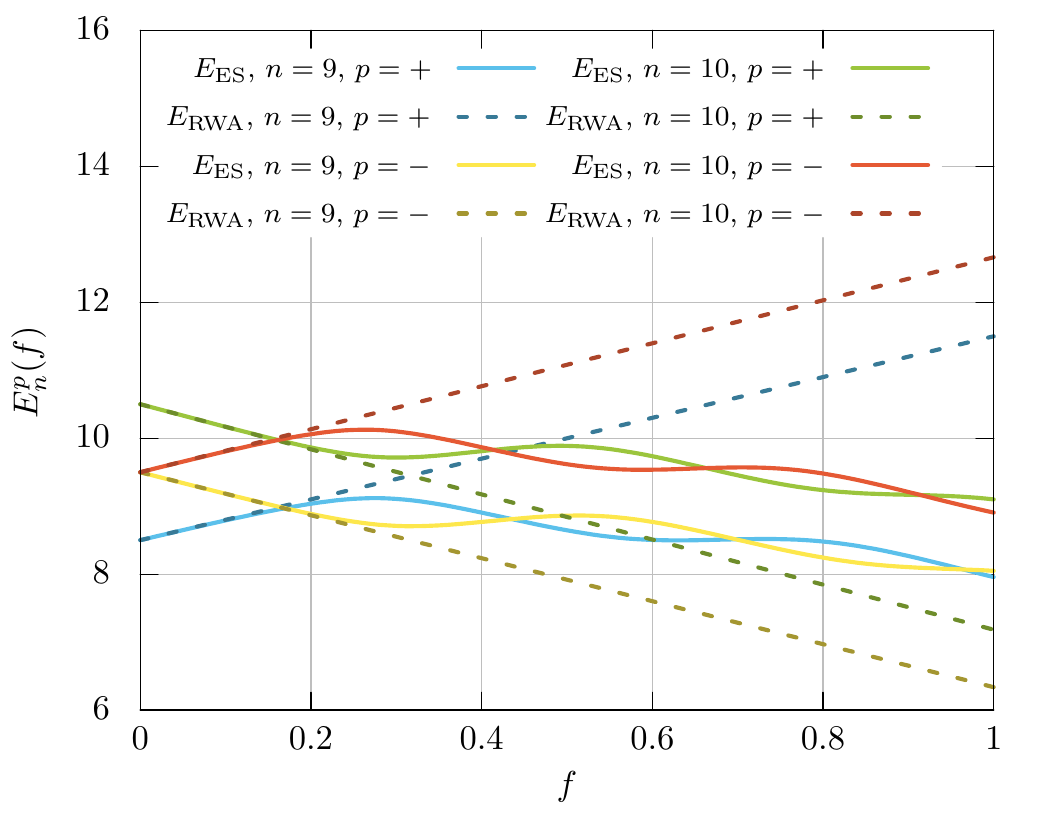}
  \caption{(Color online) The dependence of the energy levels of
    excited states $E_{n}^{p}$ on the dimensionless coupling constant
    $f$. (Left pane). The comparison of the exact numerical solution
    $E_{\mathrm{ES}}$ with the approximate analytical formulas of RWAR
    $E_{\mathrm{RWAR}}$ for the quantum Rabi model with the
    $\opA{A}^{2}$ term, i.e., $\delta = 1$. (Rigth pane). The
    comparison of the exact numerical solution $E_{\mathrm{ES}}$ with
    the approximate analytical formulas of RWA $E_{\mathrm{RWA}}$ for
    the quantum Rabi model with only $\opA{p}\cdot\opA{A}$ term, i.e.,
    $\delta = 0$. For both panes the parameter
    $\Delta = 1$.}\label{fig:3}
\end{figure*}

Operator $\opa{H}'$ depends on the two parameters $f$ and $k$ that
according to the Ref.~\cite{PhysRevLett.35.432} can not change
independently of each other if we take into account the
Thomas-Reiche-Kuhn sum rule \cite{reiche_ueber_1925} for the oscillator
strengths of the transitions of an atomic system from a state $(a)$ to
all allowed states $(b)$ with the transition frequency $\omega_{ba}$
and the dipole transition matrix element $d_{ba}$
\begin{align}
  2 m_0 \sum_b \omega_{ba} |d_{ba}|^2 = 1. \label{eq:8}
\end{align}
Indeed let us consider the quantity $f^{2}$
\begin{align}
  f^2 = \frac{2\pi e_{0}^{2}}{\omega V}\Delta^{2}d^{2} = \frac{2
  \pi e_0^2}{2 m_0 \omega^2 V}(2 m_0 \omega \Delta^2 d^2) \label{eq:9}
\end{align}
and introduce the relative oscillator strength of the resonant atomic
transition of the QRM
\begin{align}
  \frac{ \omega \Delta  d^2 }{\sum_b \omega_{b\downarrow}
  |d_{b\downarrow}|^2} \equiv \frac{1}{\delta}, \quad \delta \geq
  1. \label{eq:10}
\end{align}
Here we expressed the transition frequency as
$\omega_{\uparrow\downarrow} = \omega \Delta$. The sum in the
denominator of Eq.~(\ref{eq:8}) contains positive terms including the
term with $b = \uparrow$ \cite{PhysRevLett.35.432}. For this reason,
the quantity $\delta \geq 1$. Then from Eqs~(\ref{eq:8}) and
(\ref{eq:9}) we find the relation between the parameters $f$ and $k$
of the Hamiltonian of the QRMA
\begin{align}
  k = \frac{\delta}{\Delta} f^2, \quad \Omega = \sqrt{1 + 4
  \frac{\delta}{\Delta} f^2}. \label{eq:11}
\end{align}
We would like to stress here that the parameter $\delta \geq 1$, which
has a crucial consequence on the spectrum of the system. Exactly this
condition makes essentially more difficult for a QRMA with purely
electromagnetic interaction to reach the strong coupling regime. Due
to the $\opA{A}^{2}$ term the renormalized coupling constant has a
different scaling behavior for large values of $f$, i.e.,
$\tilde{f} = f/\sqrt{\Omega} \sim \sqrt{f}$.

The spectrum of the QRMA model as in the case of the QRM model is defined
as a solution of a system of equations
\begin{equation}
  \begin{aligned}
    \opa{H}' |\Psi_{n}^{p}\rangle
    &= E_{n}^{p}|\Psi_{n}^{p}\rangle,
    \\
    \opa{P} |\Psi_{n}^{p}\rangle &= p|\Psi_{n}^{p}\rangle
  \end{aligned}\label{eq:12}
\end{equation}
where $\opa{P} = \sigma_{3} \exp(i \pi \opa{a}^{\dag}\opa{a})$ is the
operator of a combined parity with eigenvalues $p = \pm 1$ and the
quantum number $n = 0, 1, \ldots$ numerates the field states.

The proof that the system of Eqs.~(\ref{eq:12}) is exactly integrable
is given in Ref.~\cite{PhysRevLett.107.100401}. In addition, there
exists a numerous number of works \cite{RevModPhys.91.025005} which
construct an approximate solution for the strong coupling regime or
provide the uniform approximation \cite{feranchuk_two-level_1996,
  feranchuk_strong_2011, PhysRevLett.99.173601,
  *PhysRevLett.99.259901, PhysRevA.99.033834} for a large range of
variation of the coupling constant. In order to investigate how the
QRMA behavior is modified with respect to the conventional QRM we will
compare the exact numerical solution \cite{feranchuk_physical_2016,
  feranchuk_two-level_1996} of the system of Eqs.~(\ref{eq:12}) with
the analytical solution in the framework of the rotating wave
approximation with the renormalized frequency and the coupling
constant (RWAR) \cite{lambropoulos_fundamentals_2007,
  scully_quantum_1997}.

\begin{figure*}[t]
  \centering
  \includegraphics[width=\columnwidth]{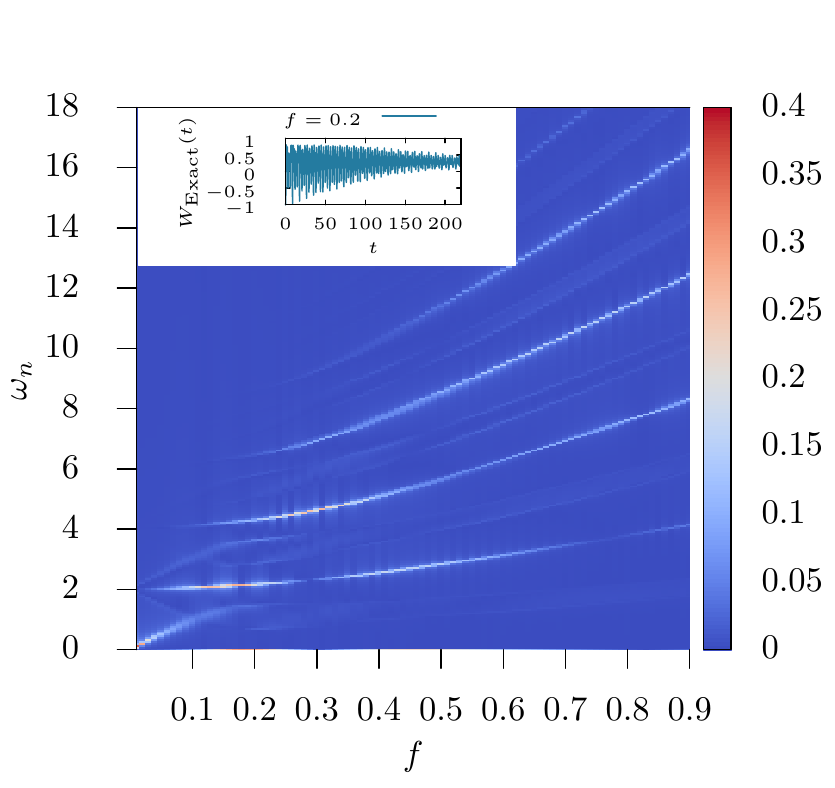}
  \includegraphics[width=\columnwidth]{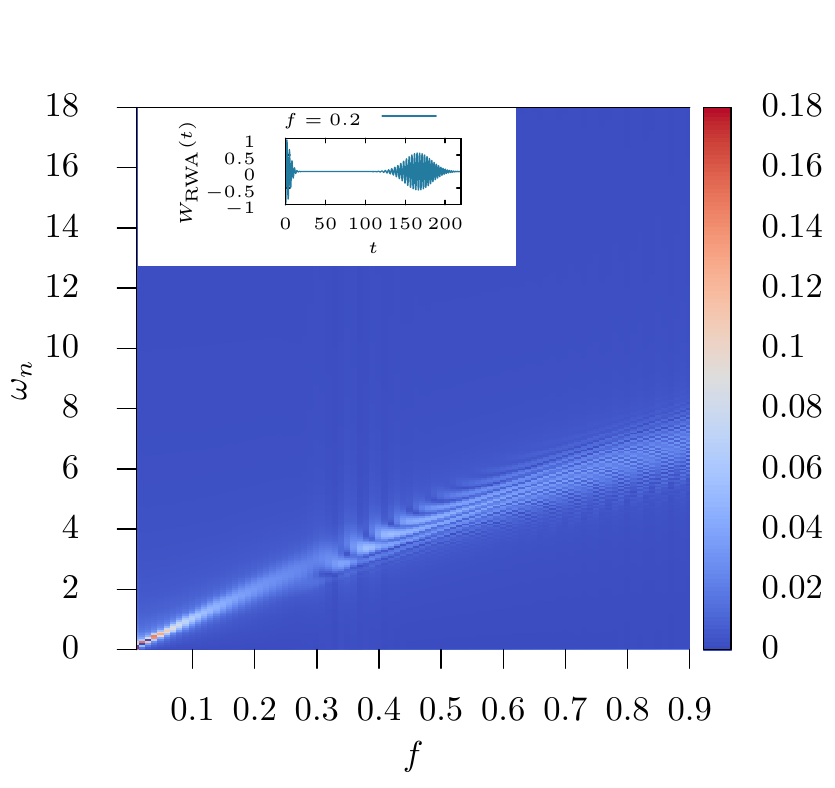}
  \caption{(Color online) Fourier transform of the inverse population
    of QRMA as a function of a dimensionless frequency $\omega_{n}$
    and a coupling constant $f$. The parameters of the system are
    $\Delta = 1$, $\delta = 1$, $\langle n\rangle = \epsilon^{2} = 25$
    and the time interval $T = 100$. (Left pane) The exact numerical
    solution of the QRMA. (Right pane) The RWA with the renormalized
    frequency and coupling constant. The inset on both planes
    represents the time evolution of the system for the value of the
    coupling constant $f = 0.2$}\label{fig:4}
\end{figure*}

For the RWAR the eigenvalues and eigenvectors are given by the well
known formulas \cite{lambropoulos_fundamentals_2007}
\begin{align}
  \begin{aligned}
    E_{\mathrm{GS}}
    &= -\frac{\Delta}{2} + \frac{\Omega - 1}{2}, \quad
    |\psi_{\mathrm{GS}}\rangle =
    \chi_{\downarrow}|0\rangle,
    \\
    E_{n}^{\pm}
    &= \Omega (n+1) -\frac{1}{2} \pm \sqrt{\frac{(\Delta -
        \Omega)^{2}}{4} + \frac{f^{2}(n+1)}{\Omega}}
    \\
    |\psi_n^{\pm}\rangle
    &= A_n^{\pm} \chi_\uparrow |n,\Omega\rangle
    + B_n^{\pm} \chi_\downarrow |n+1,\Omega\rangle,
  \end{aligned}\label{eq:13}
\end{align}
where the coefficients $A_{n}^{\pm}$ and $B_{n}^{\pm}$ are provided in
a supplementary information, $|n, \Omega\rangle = \opa{S}|n\rangle$
and $n = 0, 1, 2, \ldots$. We point out that the ground state
$E_{\mathrm{GS}}$ is obtained separately from the excited states and
can be verified by acting with $\opa{H}'$ on
$|\psi_{\mathrm{GS}}\rangle$. Moreover, the excited states are
calculated from the doubly degenerate states of the noninteracting
system.

The exact numerical solution of the system of Eqs.~(\ref{eq:12}) was
performed with the help of the Arnoldi iteration algorithm and in
addition by using an iteration scheme described in
\cite{feranchuk_two-level_1996, feranchuk_physical_2016}.

In Fig.~\ref{fig:1}, \ref{fig:3} we compare the eigenvalues as a
function of a coupling constant obtained in the framework of RWAR and
the exact numerical solution. In Fig.~\ref{fig:2} we show the behavior
of the expectation value of the photon number for the ground state of
the QRM as a function of the coupling constant. When the parameter
$\delta = 0$ the exact solution of the system in the strong coupling
regime is drastically different from the RWA, as was expected and is
well know. Consequently, the modification of the RWA is required as
was demonstrated by many works \cite{RevModPhys.91.025005}. However,
if we consider QRMA, i.e., QRM with the $\opA{A}^{2}$ term
($\delta \geq 1$), the RWA with the renormalized frequency correctly
describes the observable characteristics of QRMA for the strong
coupling regime. This statement remains correct not only for the
ground state, but also for the excited states as demonstrated in
Fig.~\ref{fig:2}.

As was demonstrated in Ref.~\cite{scully_quantum_1997} the RWA for QRM
is limited by the values of the coupling constant when the levels with
the same combined parity become degenerate for different quantum
numbers $n$, i.e., $E_{n+2}^{-}(f) = E_{n}^{-}(f)$. However, for the
QRMA due to the renormalization of the frequency and the coupling
constant this degeneracy becomes important for much larger values of
$f$, which makes RWA applicable for the whole relevant range of
variation of the coupling constant.

From the analysis of the spectrum we can conclude that the
$\opA{A}^{2}$ term qualitatively changes the system behavior. If
$\delta = 0$ the ground state energy of the quantum Rabi model is
lower then the combined energy of an atom and a field of a
noninteracting system. Therefore, if the atom is placed into the
cavity it is more preferable to form a bound state. In a stark
contrast, however, is the situation when we include the $\opA{A}^{2}$
term. In this case the ground state energy of the QRMA model is larger
than the sum of energies of noninteracting system.

Finally, in Fig.~\ref{fig:4} we study the dynamics of the QRMA
model. For this we analyzed the Fourier transform of the inverse
population $W(t)$. Despite the fact that the RWA correctly describes
the stationary states of the system it still fails to reproduce the
time dynamics. We observe that with the increase of the coupling
constant $f$ the additional frequencies appear in the spectrum in a
particular manner, namely we start observing the doubling of
frequencies that demonstrates that the system starts to exhibit a
chaotic behavior \cite{feranchuk_two-level_1996}.

A lot of works that study radiation-matter interaction are devoted to
the investigation of the counter rotating terms in the Hamiltonian of
QRM that become important in the strong coupling regime. However, as
demonstrated in this letter the behavior of the system in the strong
coupling regime is different as was previously thought. The reason for
this drastic change is the $\opA{A}^{2}$ term, which is present in the
Hamiltonian of nonrelativistic quantum electrodynamics but is often
ignored in practical applications. As a consequence, the investigation
based on the complete Hamiltonian of quantum electrodynamics
demonstrates that the QRMA is reduced to the standard QRM but with the
renormalized frequency and the coupling constant of the form
$\sim f / (1 + a f^{2})^{1/4}$ that scales as $\sqrt{f}$ for large
values of $f$ and a constant energy shift. As a result we observe the
qualitative modification of the behavior of QRMA.

\section{Supplementary information}
\textbf{Eigenvalues and eigenvectors in the rotating wave
  approximation.} The excited energy levels of the QRM in the
framework of the rotating wave approximation are given by the
following formulas \cite{lambropoulos_fundamentals_2007}
\begin{align}
  \begin{aligned}
    E_{n}^{\pm}
  &= \Omega (n+1) -\frac{1}{2} \pm \sqrt{\frac{(\Delta -
  \Omega)^{2}}{4} + \frac{4f^{2}(n+1)}{4\Omega}}
  \\
  |\psi_n^{\pm}\rangle
  &= A_n^{\pm} \chi_\uparrow |n,\Omega\rangle
  + B_n^{\pm} \chi_\downarrow |n+1,\Omega\rangle,
  \end{aligned}\label{eq:14}
\end{align}
where $n = 0, 1, 2,\ldots$,
\begin{align}
  A_n^{\pm} = \dfrac{1}{\sqrt{1 + (\lambda_n^{\pm})^2}}, \quad
  B_n^{\pm} = - \dfrac{\lambda_n^{\pm}}{\sqrt{1 +
  (\lambda_n^{\pm})^2}}, \label{eq:15}
\end{align}
and
\begin{align}
  \lambda_n^{\pm} = \dfrac{(\Delta - \Omega) \mp \sqrt{(\Delta
  -\Omega)^2 +
  4f^2(n+1)/\Omega}}{2f\sqrt{n+1}/\sqrt{\Omega}}. \label{eq:16}
\end{align}

The ground state of QRM should be investigated separately from the
excited states and reads
\begin{align}
  E_{\mathrm{GS}} = - \frac{\Delta}{2} + \frac{\Omega -1}{2}, \quad
  |\psi_{\mathrm{GS}}\rangle = \chi_\downarrow |0, \Omega
  \rangle. \label{eq:17}
\end{align}
It can be easily verified by the action of $\opa{H}$ on
$|\psi_{\mathrm{GS}}\rangle$ that $E_{\mathrm{GS}}$ is the eigenvalue.

\textbf{Numerical solution of the QRM}. In order to solve numerically
the QRM we first perform a rotation in the spin space with the
operator
\begin{align}
 \opa{R} = \frac{(1 + \ri \sigma_{2})}{\sqrt{2}} = \frac{1}{\sqrt{2}}
  \begin{pmatrix}
    1 & 1
    \\
    -1 & 1
  \end{pmatrix}. \label{eq:18}
\end{align}
With this transformation the Hamiltonian of the system and the
operator of the combined parity transform as
\begin{align}
  \opa{H}''
  &= \opa{R}^{\dag} \opa{H}' \opa{R} \nonumber
  \\
  &= \frac{\Delta}{2}
  \sigma_{1} - \frac{f}{\sqrt{\Omega}} (\opa{a} + \opa{a}^{\dag})
    \sigma_{3} + \Omega \opa{a}^{\dag}\opa{a} + \frac{\Omega -
    1}{2}, \label{eq:19}
  \\
  \opa{P}''
  &= \opa{R}^{\dag} \opa{P} \opa{R} = \sigma_{1}\exp(i \pi
  \opa{a}^{\dag}\opa{a}). \label{eq:20}
\end{align}
By writing the wave function in matrix form
$|\psi\rangle = {{u}\choose{v}}$, the operator of the combined parity
allows one to express the lower component via the upper one as $v =
p\exp(i \pi \opa{a}^{\dag}\opa{a}) u$. After the substitution of $v$
into the matrix equations we arrive to the Schr\"{o}dinger equation
for the one component wave function $u$
\begin{align}
  \Bigg(\Omega \opa{a}^{\dag}\opa{a} + \frac{\Omega - 1}{2} +
  \frac{\Delta}{2} p e^{i \pi \opa{a}^{\dag}\opa{a}} -
  \frac{f}{\sqrt{\Omega}}(\opa{a} + \opa{a}^{\dag})\Bigg)u = E
  u. \label{eq:21}
\end{align}

After this we expand the wave function $u$ in the Harmonic oscillator
basis $u = \sum_{l = 0}^{\infty}C_{l}|l\rangle$ and numerically
diagonalize the matrix
\begin{align}
  H_{kn}
  &= \left(\Omega n + \frac{\Omega - 1}{2} + \frac{\Delta}{2} p
  (-1)^{n}\right)\delta_{kn} \nonumber
  \\
  &\mspace{40mu}- \frac{f}{\sqrt{\Omega}}(\sqrt{n}\delta_{kn-1} +
  \sqrt{n+1}\delta_{kn+1}).\label{eq:22}
\end{align}

As a result the normalized state vectors of the system are given by
\begin{align}
  |\psi_{n}^{p}\rangle = \sum_{l = 0}^{\infty}
  C_{l}^{np}\frac{1}{2}{{(-1)^{l}p + 1}\choose{(-1)^{l}p -
  1}}\opa{S}|l\rangle, \label{eq:23}
\end{align}
where $\vec{C}^{np} = \{C_{l}^{np}\}$ are the normalized eigenvectors
of the matrix $H_{kn}$.

\textbf{The average number of photons}. If the QRMA is prepared in the
state with $0$ photons, then the average number of photons in the RWA
is given by
\begin{align}
  \langle \opa{n}\rangle_{\mathrm{RWA}} = \langle
  \psi_{\mathrm{GS}}|\opa{S} \opa{a}^{\dag} \opa{a}
  \opa{S}^{\dag}|\psi_{\mathrm{GS}}\rangle = \frac{(\Omega -
  1)^{2}}{4\Omega}. \label{eq:24}
\end{align}

The average number of photons for the exact solution reads
\begin{align}
  \langle \opa{n}\rangle_{\mathrm{ES}}
  &= \frac{(\Omega - 1)^{2}}{4\Omega} + \frac{\Omega^{2} + 1}{2\Omega}
    \sum _{k} k |C_{k}^{np}|^{2} \nonumber
  \\
  &+ \frac{1}{4}\left(\frac{1}{\Omega} -
    \Omega\right)\sum_{k}\sqrt{(k+1)(k+2)} \nonumber
  \\
  &\mspace{100mu}\times(C_{k}^{np*}C_{k+2}^{np} +
  C_{k+2}^{np*} C_{k}^{np}). \label{eq:25}
\end{align}

\textbf{The evolution of the QRM}. We consider that an atom in the
initial moment of time was in the lower state $\chi_{\downarrow}$ and
the field was prepared in the coherent state with the amplitude
$\epsilon$
\begin{align}
  |\Psi(0)\rangle
  &= \chi_{\downarrow}|\epsilon\rangle = \chi_{\downarrow}
    e^{\epsilon(\opa{a}^{\dag} - \opa a)}|0\rangle\label{eq:26}
\end{align}
and will characterize the system dynamics with the inverse population
\cite{scully_quantum_1997}, which in the RWA is given by the formula
\begin{align}
  W(t)
  &= \sum_{n = 0}^{\infty}
    \frac{e^{-\tilde{\epsilon}^2}\tilde{\epsilon}^{2n}}{n!
    \omega_{A}^{2}(n)}\Bigg((\Delta - \Omega)^{2} \nonumber
  \\
  &\mspace{130mu} + \frac{4f^{2}(n+1)}{\Omega}
    \cos[\omega_{A}(n)t]\Bigg), \label{eq:27}
\end{align}
where
\begin{align}
  \omega_A (n)
  &= \sqrt{(\Delta -\Omega)^2+\frac{4f^2(n+1)}{\Omega}} \nonumber
\end{align}
and $\tilde{\epsilon} = \epsilon(\Omega + 1) / (2 \sqrt{\Omega})$.

In order, to obtain the exact numerical solution we proceed in the
following way. The time dependent wave function is represented as an
expansion over stationary states of the QRM
\begin{align}
  |\Psi(t)\rangle = \sum_{np}A_{np} |\psi_{n}^{p}\rangle e^{-i
  E_{n}^{p} t}, \label{eq:28}
\end{align}
where the coefficients $A_{np}$ are determined from the initial
condition (\ref{eq:26}), i.e.,
\begin{align}
  {{0}\choose{1}}\langle k|\opa{S}^{\dag}|\epsilon\rangle = \sum_{np}
  A_{np} C^{np}_{k} \frac{1}{2} {{(-1)^{k} + 1}\choose{(-1)^{k} -
  1}}. \label{eq:29}
\end{align}
For this we rewrite the system of equations for the coefficients
$A_{np}$ in matrix form
\begin{align}
  \begin{pmatrix}
    \vec A_{+} & \vec A_{-}
  \end{pmatrix}
  &\begin{pmatrix}
    C^{+} & 0
    \\
    0 & C^{-}
  \end{pmatrix}\begin{pmatrix}
    D\left(\frac{(-1)^{k} + 1}{2}\right) & D\left(\frac{(-1)^{k} -
        1}{2}\right)
    \\
    D\left(\frac{(-1)^{k+1} + 1}{2}\right) & D\left(\frac{(-1)^{k+1} -
        1}{2}\right)
  \end{pmatrix} \nonumber
  \\
  &= \begin{pmatrix}
    0 & \{\langle k|\opa{S}^{\dag}|\epsilon\rangle\}
  \end{pmatrix},\label{eq:30}
\end{align}
where the sign $+$ or $-$ denotes $p = +1$ and $p = -1$
correspondingly, $\vec A_{\pm}$ are the vectors of unknowns of the
size $1\times N$, $C^{\pm}$ are the matrices of eigenvectors of the
size $N \times N$ (the first row is the first eigenvector, the second
row in the second one, ...), the notation
$D\left(((-1)^{k}+1) / 2\right)$ denotes the diagonal matrix, where
the diagonal is formed by the sequence $((-1)^{k}+1) / 2$ with
$k = 0, 1, \ldots$ and
$\{\langle k|\opa{S}^{\dag}|\epsilon\rangle\} = \{\langle
0|\opa{S}^{\dag}|\epsilon\rangle, \langle
1|\opa{S}^{\dag}|\epsilon\rangle, \ldots\}$.

After the coefficients $A_{np}$ are determined we computed the wave
function of the system
\begin{align}
  &\begin{pmatrix}
    \vec A_{+} & \vec A_{-}
  \end{pmatrix}
  \begin{pmatrix}
    C^{+} & 0
    \\
    0 & C^{-}
  \end{pmatrix}\begin{pmatrix}
    D\left(e^{-i E_{n}^{+} t}\right) & 0
    \\
    0 & D\left(e^{-i E_{n}^{-} t}\right)
  \end{pmatrix} \nonumber
  \\
  &\times\begin{pmatrix}
    D\left(\frac{(-1)^{k} + 1}{2}\right) & D\left(\frac{(-1)^{k} -
        1}{2}\right)
    \\
    D\left(\frac{(-1)^{k+1} + 1}{2}\right) & D\left(\frac{(-1)^{k+1} -
        1}{2}\right)
  \end{pmatrix} \nonumber
  \\
  &= \begin{pmatrix}
    \vec \psi_{+} & \vec \psi_{-}
  \end{pmatrix}.\label{eq:31}
\end{align}

As a result the density matrix of the atomic system is determined
\begin{align}
  \rho_{\mathrm{A}} = \mathrm{Sp}_{\mathrm{F}} (|\Psi(t)\rangle
  \langle\Psi(t)|) =
  \begin{pmatrix}
    \vec\psi_{+}^{\dag}\cdot\vec \psi_{+} &
    \vec\psi_{+}^{\dag}\cdot\vec \psi_{-}
    \\
    \vec\psi_{-}^{\dag}\cdot\vec \psi_{+} &
    \vec\psi_{-}^{\dag}\cdot\vec \psi_{-}
  \end{pmatrix}. \label{eq:32}
\end{align}
Finally, the inverse population is expressed through the atomic
density matrix as
\begin{align}
  W(t) = \mathrm{Sp}(\rho_{\mathrm{A}}\sigma_{3}).\label{eq:33}
\end{align}

\bibliography{biblnrwa}

\end{document}